\documentclass{PoS}

\usepackage{cite}
\usepackage{multicol}
\usepackage[titletoc]{appendix}
\usepackage[authoryear]{natbib}
\bibpunct{(}{)}{;}{a}{}{,}

\makeatletter

\newcommand{\Rmnum}[1]{\expandafter\@slowromancap\romannumeral #1@}
\makeatother

\title{Giant radio galaxies as probes of the ambient WHIM in the era of the SKA}

\ShortTitle{GRGs as probes of ambient WHIM}

\author{
		\speaker{Bo Peng}$^{1,2}$,Ru-Rong Chen$^{1}$, Richard Strom$^{1,3}$
		\\
        $^1\rm NAOC$;
        $^2\rm JLRAT$;
        $^3\rm ASTRON$
        \\
        E-mail: \email{pb@bao.ac.cn}
        }

\abstract{The missing baryons are usually thought to reside in galaxy filaments as warm-hot intergalactic medium (WHIM). From previous studies, giant radio galaxies are usually associated with galaxy groups, which normally trace the WHIM. We propose observations with the powerful SKA1 to make a census of giant radio galaxies in the southern hemisphere, which will probe the ambient WHIM. The radio galaxies discovered will also be investigated to search for dying radio sources. With the highly improved sensitivity and resolution of SKA1, more than 6,000 giant radio sources will be discovered within 250 hours.}

\FullConference{Advancing Astrophysics with the Square Kilometre Array,\\
		June 8-13, 2014\\		
		Giardini Naxos, Sicily, Italy
		}

\begin{document}

\section{The WHIM}

Baryonic matter - protons, neutrons, and electrons - belongs to a class of massive elementary particles which make up the atoms of all materials on Earth and in the stars. From the Big Bang theory and the results of WMAP, baryonic matter contributes no more than 5 percent of the cosmos, though it remains not completely accounted for. From a census of the observable baryonic density in the current Universe (e.g. \cite{shull12}), all the stars, dust and gas within galaxies constitute less than 50 percent of baryons predicted by the theory. By analyzing light from distant quasars, the number of baryons in cosmic history can be counted. Its amount has remained constant even going back to 10 billion years ago. Quantifying the missing baryons is essential to obtaining a complete picture of galaxy formation and evolution. The missing baryons are generally thought to reside between galaxies, as warm-hot intergalactic medium (WHIM) with temperatures of $10^5 - 10^7$ K. Material in the WHIM is highly ionized and can only be observed at far-ultraviolet or low-energy X-ray wavelengths. This diffuse plasma is a common prediction in the $\Lambda$ cold dark matter ($\Lambda$CDM) cosmologies (e.g. \cite{dave10} ). However, because its density is relatively low, the detection remains a challenge \citep{dave01, wern08, dietrich12}.

\section{ Giant radio galaxies (GRGs)}

Giant radio galaxies (GRGs) (linear size > 0.7 Mpc, h = 0.71) are a quite unique category, as they are the biggest objects in the sky with large radio structures. It is generally believed that their radio lobes are powered by beams of relativistic particles and magnetic fields from an active nucleus. Radio structures of many known GRGs are asymmetric over their mega-parsec extents \citep{schoen00, lara01, machal01, sari05, strom13} (see Figure 1). As the radio lobes expand, they interact with the ambient medium and can act as tracers of this barely detectable material. The large extent of GRGs provides a good opportunity to investigate the ambient gas.

\begin{figure}
\centering
\includegraphics[angle=-90, width=10cm]{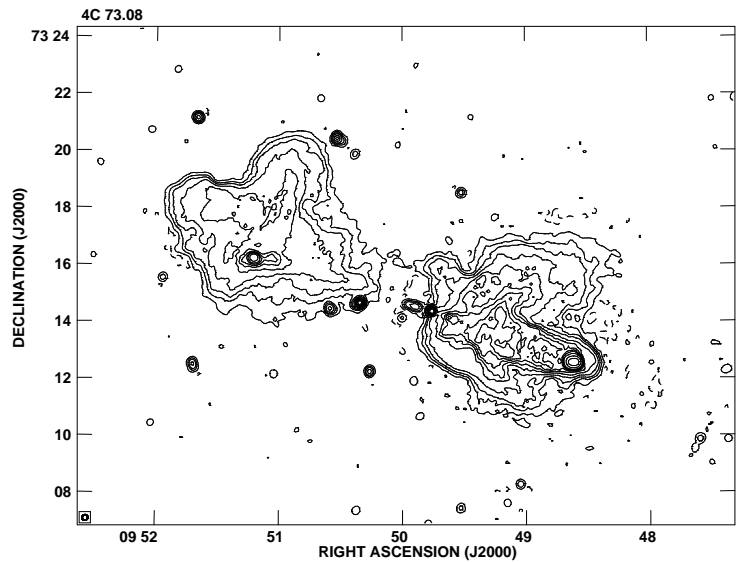}
\caption{WSRT map at 21 cm wavelength of the radio galaxy 4C~73.08. The brighter western lobe is more compact than its eastern counterpart. A similar effect is seen in NGC~6251 (Figure.2).}
\end{figure}

\begin{figure}
\centering
\includegraphics[angle=-90,width=10cm]{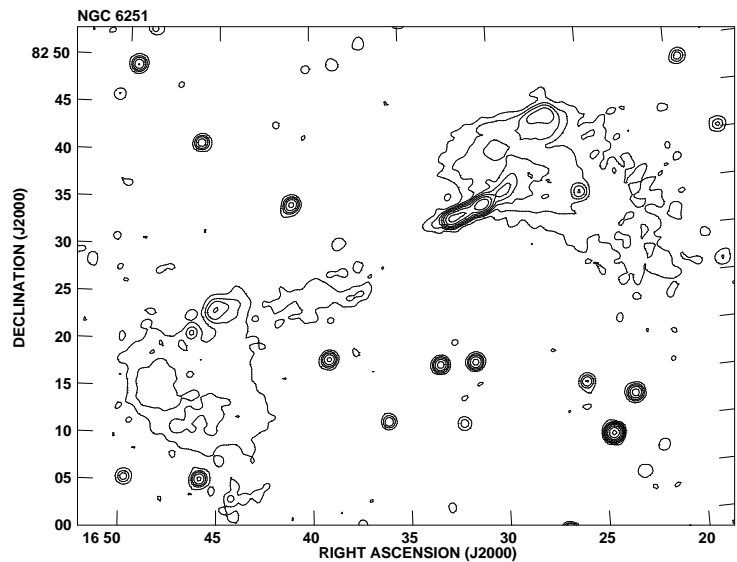}
\caption{WSRT map made at 92 cm wavelength of the radio galaxy NGC~6251. Like all of the GRGs, it is srongly polarized at long wavelengths. Note how unequal the distances between the ends of the two lobes and the host are.}
\end{figure}

\begin{figure}
\centering
\includegraphics[width=10cm]{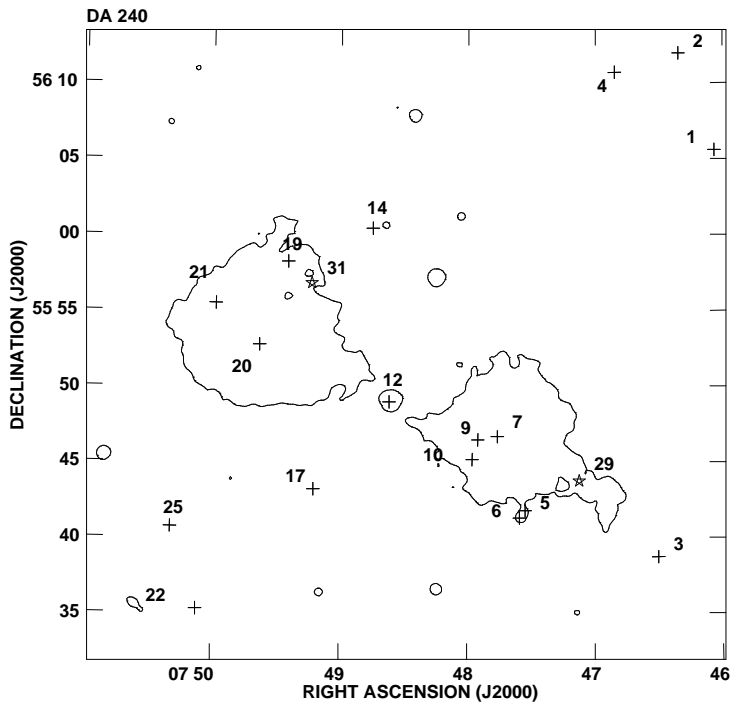}
\caption{WSRT map made at 49 cm wavelength of the radio galaxy DA~240. Galaxies which are members of the DA~240 group have their positions indicated by symbols.}
\end{figure}

\section{Radio galaxies and cosmology}
By 1950, several compact radio sources had been associated with distant galaxies \citep{bolton49, ryle50}, and with the identification of the very strong radio source Cygnus A \citep{smith51, baade54} with a $z = 0.056$ galaxy, the potential of radio sources for probing the Universe became apparent. This expectation was confirmed when 3C 295 was shown to be associated with a $z = 0.461$ galaxy \citep{min60}. Simultaneously, radio source structure was found to be double lobed \citep{jennison53}, with sizes ranging from parsecs to hundreds of kilo-parsecs.

Much effort went into identifying (and measuring the redshifts of) radio sources from reliable catalogues, especially 3C. High resolution radio maps (made by the One Mile Telescope, Westerbork and the VLA) tied down the radio source structures. The angular size measured could be translated into linear dimensions. Inverse Compton scattering of the radio emitting high-energy electrons becomes increasingly important at greater redshift. This effect could "snuff out" radio sources at high $z$ \citep{miley68} before they could attain the dimensions found among the largest radio galaxies in the local Universe.

\section{Tracing the WHIM with GRGs}
As was clear from early WSRT observations of DA~240, 3C~236 and NGC~6251 (Figure 2), giant radio galaxies are strongly polarized at 608 MHz \citep{willis74}. The implied low ambient densities were at least qualitatively consistent with the large linear sizes of the radio components. A definitive study of five GRGs carried out by \cite{mack97} mapped their linear polarization at wavelengths from 92 to 2.8 cm. All of the sources were found to be polarized in many places at the longest wavelengths, indicating an internal rotation measure of $\sim$ 1 rad m$^{-2}$. The implied densities within the lobes depend upon the magnetic field strength. For equipartition field strengths of typically 1 {$\mu$}G, the particle densities are of order 10$^{-5}$ cm$^{-3}$. In a more general consideration of densities associated with GRGs, \cite{mack98} conclude that the IGM with which the radio components interact has a density of 1 - 4$\times$ 10$^{-5}$ cm$^{-3}$.

Several recent studies have explored the use of GRGs for probing the WHIM. In both the cases of \cite{sub08}, and \cite{safour09}, sensitive radio continuum images of two GRGs were used to relate their morphology to the ambient medium on the Mpc scale. They made spectroscopic observations to investigate galaxy distribution within a 2$^{\circ}$ field surrounding the radio galaxy hosts. Both studies showed clear correspondence between the giant synchrotron lobes and the surrounding galaxy distributions. Such a correspondence was also found in a series of papers \citep{chen11b, chen12a, chen12b}. They found the existence of a galaxy group around each giant radio galaxy (NGC 6251, NGC~315 and 4C~73.08) through optical spectroscopic observations. However, in the case of DA~240 \citep{peng04, chen11a}, the relatively symmetrical lobes coincide with a galaxy group. The group members evenly reside along the major axis of the lobes while the isolated host galaxy lies in the middle (Figure 3). The expected X-ray luminosities are much higher than the observed values, which suggests that the matter density within the lobes is quite low. A similar conclusion was obtained by \cite{mala13}. In the investigation of a sample of 12 GRGs, they derived lobe energy densities from the radio observations via equipartition arguments. The inferred pressures in the lobes of the giant radio sources, which range from 1.1$\times$ 10$^{-15}$ to 2.0$\times$ 10$^{-14}$ Pa, are lower than inferred from X-ray observations of dense filaments. This demonstrates the potential of using GRGs to study the WHIM gas within which their lobes evolve and interact.

\section{Prospects with SKA1}
A significant fraction of known giant radio galaxies were identified from the surveys of the WSRT and VLA: WENSS, NVSS and FIRST. The SKA1\_SUR will be equipped with 96 dishes in Western Australia, including 36 dishes of ASKAP (latitude is about 27$^{\circ}$S). At L band, it will have a high survey speed figure of merit (FoM) of about 156 times greater than the VLA, and 1.5 times better sensitivity as well as a 1.6 to 49 times higher angular resolution (Table 1). If it carries out a survey similar to the NVSS, with integration times of 10 minutes and a bandwidth of 500 MHz at L band, the noise level should achieve 9.1 {$\mu$}Jy/beam. Since the GRGs usually have intermediate power between 10$^{25}$ and 10$^{26}$ W/Hz at 1.4 GHz, if the $SNR$ is presumed (based on experience) to be 150 \citep{schoen01, machal01}, the luminosity distance will be 7825 Mpc ($z \sim 1.2$). Since the minimum elevation angle is about 15 degree (similar to the ASKAP),  SKA1\_SUR can observe the sky up to +\,48$^{\circ}$ (more than 80 percent of the whole sky). If the Galactic latitude is set to be $\vert{b}\vert > 12.5^{\circ}$ to avoid the confusion of the Milky Way, the sky coverage will then be reduced to about 27,000 deg$^2$. Adopting a number density of 10$^{-7}$ Mpc$^{-3}$ \citep{sari05}, the SKA1\_SUR is expected to detect about 12,000 GRGs in a total observing time of about 250 hours (formula listed in the Appendix). Even at an early phase (50\% capability), SKA1\_SUR will still be able to detect about 6,000 GRGs, which can be identified with counterparts obtained from optical surveys, such as SDSS, 2dF, and the planned LSST. The GRG sample will then increase by more than an order of magnitude.

At the same time, the SKA1\_MID will be equipped with 254 dishes in South Africa, including 64 dishes from MeerKAT. Compared with the VLA, this will provide 6.2 times better sensitivity and 74 times higher survey speed. Moreover, the angular resolution will be 6.4 to 200 times better. After GRG candidates have been obtained from the SKA1\_SUR, SKA1\_MID can be used for a deeper investigation, especially for low surface brightness lobes, which is important for their use as cosmological probes. If the SKA2\_MID could be equipped with PAFs, the survey speed with SKA2\_MID would be greatly increased, and this will make the SKA2\_MID quite a powerful instrument for discovering more of these faint, extended radio sources at higher redshift.

Furthermore, since the frequency $\nu$ of radio synchrotron emission is proportional to the square of the electrons' energy $E$ and the ambient magnetic field $B$ ($\nu \sim E^{2}B$), which is expected to be low and weak in the diffuse radio lobes of GRGs, detection of new radio structures in GRGs is likely from observations using the SKA1 low frequency array. There may even be discovery of new ones (like the discovery of GRG UGC 09555 with LOFAR). The low frequency array of SKA1 will be equipped with 250,000 antennas in Western Australia, with 16 times higher sensitivity and 520 times quicker survey speed compared with LOFAR.

Recently, \cite{luo10} modelled the evolution of FR\small{\Rmnum{1}} and \small{\Rmnum{2}} radio sources, finding that the latter evolve with decreasing monochromatic radio power, while the power of the former increases with age. The final stage for both sees rapidly declining emission as the radio source fades away, the lobes receiving little or no energy from the host nucleus or jets. As the lobes decline, their radio spectra will steepen, so a dying radio galaxy should have lobes with a steep radio spectrum and no jets or nuclear component. Finding candidates for this GRG phase will be well-suited to the SKA1 capability at low frequencies.

With the highly improved system sensitivity, survey speed and angular resolution of SKA1, there should be more than one order of magnitude increase in the number of GRGs from about 140 at present (a small fraction of which have redshifts greater than 0.5 \citep{komb09}). With their ability to trace the ambient IGM, more GRGs will help illuminate the distribution and physical state of the WHIM and identify whether it can explain the missing baryon problem. And with more details on high-$z$ GRGs, being able to compare them with low redshift sources will help to clarify how giant radio galaxies evolve, and especially how their environments affect their radio structures, and whether there are significant differences in group size, composition, morphology, etc., with cosmic epoch.

At low redshifts, SKA1 with its excellent sensitivity and angular resolution will be able to investigate GRG groups in unprecedented detail. In several groups (like DA~240, \cite{peng04,chen11a}; and 4C~73.08, \cite{strom13, chen12b}) we find that a large fraction of the companion galaxies are weak radio sources. The SKA1 sensitivity will enable us to explore the weak radio emissions of companion galaxies further. In addition, a number of the group galaxies appear to be spiral galaxies. 21 cm H{\small\Rmnum{1}} observations of them are likely to turn up a number of detections. One can use the H{\small\Rmnum{1}} data to estimate the amount of mass present, and study the group kinematics.

\begin{table}
\caption{Summary of system sensitivity, survey speed and angular resolution of SKA1, VLA, WSRT at L band, and LOFAR at low frequency (extracted from \cite{dew13}).}
\begin{tabular}{p{40pt}p{40pt}cccccc}
\hline
	&	&\small JVLA	&\small WSRT	&\small SKA1\_MID	&\small SKA1\_SUR	&\small LOFAR	&\small SKA1\_LOW\\
\hline
\small$\rm A_{sys}/T_{sys}$	& \small $\rm m^{2}/K$	& 265	&124	&	1630	&391	&61	&1000\\
\tiny Survey\,Speed\,FoM &\small	$\rm deg^{2}m^{4}K^{-2}$& 1.76E4	&3.84E3	&1.3E6	&2.75E6	&5.21E4	&2.7E7\\
\small Resolution &\small	arcsec&	1.4 - 44	&16		&0.22	&0.9	&5	&11\\
\small FoV	&\small deg$^2$	&0.25	&0.25	&0.49	&18	&14	&27	\\
\small Bandwidth	&\small MHz	&1000	&160	&250	&500	&4	&250\\
\hline
\end{tabular}

\end{table}

\section{Conclusions}
Our goal with SKA1 is to carry out a large-scale survey to search for giant radio galaxies. The survey should be unbiased and complete down to a well-defined brightness or flux density, over a specified sky area. The GRGs found will be used to trace out the WHIM, while we also hope to detect candidate dying RGs, objects of low surface brightness with little evidence for compact features (jets, nuclear components), but with a steep radio spectrum. The high-$z$ GRGs will be used to investigate how giant radio sources evolve. Some of the relatively nearby groups (like that surrounding DA~240) appear to have several spiral companions. We plan to observe them in the 21 cm H{\small\Rmnum{1}} line to study the motion and masses of GRG group members.

Among the questions we hope to answer, we note the following: Can one use the GRGs to detect the WHIM? And if so, can we pinpoint the missing baryons? Can one use GRGs and their surrounding companions to trace the distribution and physical state of the WHIM? Finally, are there enough high-$z$ giant radio galaxies to investigate how radio galaxies evolve?

\section*{Acknowledgements}
This work was partially supported by the Chinese Ministry of Science and Technology under the State Key Development Program for Basic Research, Grant Nos. 2013CB837900 and 2012CB821800, the Projects of International Cooperation and Exchange, NSFC, Grant No. 11261140641, the key research program of the Chinese Academy of Sciences (CAS), Grant No. KJZD-EW-T01, and the Open Project Program of the Key Laboratory of Radio Astronomy.  RGS wishes to thank the Chinese Academy of Sciences for the award of a visiting professorship, when most of this research was done.

\begin{multicols}{2}
\bibliographystyle{apj}

\begin{thebibliography}{99}
\footnotesize{
\bibitem[\protect\citeauthoryear{Baade \&\ Minkowski}{1954}]{baade54}Baade, W. \&\ Minkowski, R., 1954, ApJ, 119, 215
\bibitem[\protect\citeauthoryear{Bolton, Stanley \&\ Slee}{1949}]{bolton49}Bolton, J., Stanley, G., \&\ Slee, O., 1949, Nature, 164, 101
\bibitem[\protect\citeauthoryear{Burns}{1998}]{burns98}Burns, J., 1998, Sci, 280, 400
\bibitem[\protect\citeauthoryear{Chen et al.} {2011a}]{chen11a}Chen, R., Peng, B., Strom, R., Wei, J., \&\ Zhao, Y.,  2011a, A{\&}A, 529, A5
\bibitem[\protect\citeauthoryear{Chen et al.}{2011b}]{chen11b}Chen, R., Peng, B., Strom, R., \&\ Wei, J., 2011b, MNRAS, 412, 2433
\bibitem[\protect\citeauthoryear{Chen et al.}{2012a}]{chen12a}Chen, R., Peng, B., Strom, R., \&\ Wei, J., 2012a, MNRAS, 420, 2715
\bibitem[\protect\citeauthoryear{Chen et al.}{2012b}]{chen12b}Chen, R., Peng, B., Strom, R., \&\ Wei, J., 2012b, MNRAS, 422, 3004
\bibitem[\protect\citeauthoryear{de Zotti et al.}{2010}]{zotti10}de Zotti, G., Massardi, M., Negrello, M., \&\ Wall, J., 2010, A{\&}ARv, 18, 1
\bibitem[\protect\citeauthoryear{Dav\'e et al.}{2001}]{dave01}Dav\'e, R., Cen, R., Ostriker, J., Bryan, G., Hernquist, L. et al., 2001, ApJ, 552, 473
\bibitem[\protect\citeauthoryear{Dav\'e et al.}{2010}]{dave10}Dav\'e, R., Oppenheimer, B., Katz, N., Kollmeier, J., \&\ Weinberg, D., 2010, MNRAS, 408, 2051
\bibitem[\protect\citeauthoryear{Dewdney et al.}{2013}]{dew13}Dewdney, P., Turner, W., Millenaar, R., MacCool, R., Lazio, J., Cornwell, T., 2013, "SKA1 System Baseline Design", Document number SKA-TEL-SKO-DD-001 Revision 1
\bibitem[\protect\citeauthoryear{Dietrich et al.}{2012}]{dietrich12}Dietrich, J., Werner, N., Clowe, D., Finoguenov, A., Kitching, T. et al., 2012, Nat, 487, 202
\bibitem[\protect\citeauthoryear{Jennison \&\ Das Gupta}{1953}]{jennison53}Jennison, R. \&\ Das Gupta, M., 1953, Nat, 172, 996
\bibitem[\protect\citeauthoryear{Komverg \&\ Pashchenko}{2009}]{komb09}Komberg, B. \&\ Pashchenko, I., 2009, ARep, 52, 1086
\bibitem[\protect\citeauthoryear{Lara et al.}{2001}]{lara01}Lara, L., Cotton, W., Feretti, L., Giovannini, G., Marcaide, J. et al., 2001, A{\&}A, 370, 409
\bibitem[\protect\citeauthoryear{Luo \&\ Sadler}{2010}]{luo10}Luo, Q. \&\ Sadler, E. M., 2010, ApJ, 713, 398
\bibitem[\protect\citeauthoryear{Mack et al.}{1997}]{mack97} Mack, K., Klein, U., O' Dea, C. P., \&\ Willis, A. G. 1997, A\&A, 123, 423
\bibitem[\protect\citeauthoryear{Mack et al.}{1998}] {mack98} Mack, K., Klein, U., O' Dea, P., Willis, G., \&\ Saripalli, L. 1998, A\&A, 329, 431

\bibitem[\protect\citeauthoryear{McCarthy, van Breugel \&\ Kapahi}{1991}]{mccar91}McCarthy, P., van Breugel, W. \&\ Kapahi, V., 1991, ApJ, 371, 478
\bibitem[\protect\citeauthoryear{Machalski, Jamrozy \&\ Zola}{2001}]{machal01}Machalski, J., Jamrozy, M., Zola, S., 2001, A\&A, 371, 445
\bibitem[\protect\citeauthoryear{Malarechi et al.}{2013}]{mala13}Malarecki, J., Staveley-Smith, L., Saripalli, L., Subrahmanyan, R., Jones, D. et al., 2013, MNRAS, 432, 200
\bibitem[\protect\citeauthoryear{Miley}{1968}]{miley68}Miley, G., 1968, Nat, 218, 933
\bibitem[\protect\citeauthoryear{Minkowski}{1960}]{min60}Minkowski, R., 1960, ApJ, 132, 908
\bibitem[\protect\citeauthoryear{Peng et al.}{2004}]{peng04}Peng, B., Strom, R., Wei, J., \&\ Zhao, Y., 2004, A{\&}A, 415, 487
\bibitem[\protect\citeauthoryear{Ryle, Smith \&\ Elsmore}{1950}]{ryle50}Ryle, M., Smith, F., \&\ Elsmore, B., 1950, MNRAS, 110, 508
\bibitem[\protect\citeauthoryear{Safouris et al.}{2009}]{safour09}Safouris, V., Subrahmanyan, R., Bicknell, G., \&\ Saripalli, L., 2009, MNRAS,393, 2
\bibitem[\protect\citeauthoryear{Saripalli et al.}{2005}]{sari05}Saripalli, L., Hunstead, R., Subrahmanyan, R., \&\ Boyce, E., 2005, AJ, 130,896
\bibitem[\protect\citeauthoryear{Schoenmakers et al.}{2000}]{schoen00}Schoenmakers, A., Mack, K.-H, de Bruyn, A., Rottgering, H., Klein, U., \&\ van der Laan, H., 2000, A{\&}AS, 146, 293
\bibitem[\protect\citeauthoryear{Schoenmakers et al.}{2000}]{schoen01}Schoenmakers, A., de Bruyn, A., R$\ddot{\rm o}$ttgering, H. \&\ van der Laan, H. 2001, A\&A, 374, 861
\bibitem[\protect\citeauthoryear{Shull, Smith \&\ Danforth}{2012}]{shull12}Shull, J., Smith, B. \&\ Danforth, C., 2012, ApJ, 759, 23
\bibitem[\protect\citeauthoryear{Smith}{1951}]{smith51}Smith, F., 1951, Nat, 168, 555
\bibitem[\protect\citeauthoryear{Strom et al.}{2013}]{strom13}Strom, R., Chen, R., Yang, J., \&\ Peng, B., 2013, MNRAS, 430, 2090
\bibitem[\protect\citeauthoryear{Subrahmanyan et al.}{2008}]{sub08}Subrahmanyan, R., Saripalli, L., Safouris, V., \&\ Hunstead, R., 2008, ApJ,677, 63
\bibitem[\protect\citeauthoryear{Werner et al.}{2008}]{wern08}Werner, N., Finoguenov, A., Kaastra, J., Simionescu, A., Dietrich, J. et al., 2008, A{\&}A, 482, L29
\bibitem[\protect\citeauthoryear{Willis, Strom \&\ Wilson}{1974}]{willis74}Willis, A.G., Strom, R.G. \&\ Wilson, A.S., 1974, Nat, 250, 625
}
\end{thebibliography}


\section*{Appendix}
\begin{appendices}
\footnotesize{
\section{Formula used to estimate the number of GRGs for SKA\_SUR}

Number of giant radio galaxies: 
\begin{equation}
N=\rho \times V_{\rm co}
\end{equation}

Number density: $\rho = 10^{-7}$ ($\rm Mpc^{-3}$)

Comoving volume: ($\rm Mpc^{3}$)
\begin{equation}
V_{\rm co}=\frac{1}{3}A_{\rm sky}D_{\rm co}^{3}
\end{equation}

Sky coverage: (deg$^2$)
\begin{equation}
A_{\rm sky}= \Delta{RA}\times\Delta{\rm{sin}(\textit{Dec})}-\Delta{l}\times\Delta{\rm{sin}(\textit{b})}
\end{equation}

Comoving distance: (Mpc)
\begin{equation}
D_{\rm co}=\frac{D_{ L}}{1+z}
\end{equation}

Redshift $z$:\ (www.icosmos.co.uk)

Luminosity distance: (Mpc)
\begin{equation}
D_{ L}=\sqrt{\frac{L_{\nu}}{4{\pi}S_{\rm obs}}}
\end{equation}

Luminosity at frequency of $\nu$:\ $L_{\nu}$, $10^{25} - 10^{26}$ W/Hz at 1.4 GHz

Flux density observed with telescope: (Jy)
\begin{equation}
S_{\rm obs}=SNR\times\sigma
\end{equation}

Sigma to noise: $SNR > 150$

Noise level with two polarizations: (Jy/beam)
\begin{equation}
\sigma=\frac{S_{\rm sys}}{\sqrt{2B\tau}}
\end{equation} 

System equivalent flux density:
\begin{equation}
S_{\rm sys}=\frac{2k_{\rm b}T_{\rm sys}}{A_{\rm eff}}
\end{equation}

Boltzman constant: $k_{\rm b} = 1.38\times 10^{-23}$ J/K

For SKA1\_SUR, sensitivity specification: $\frac{A_{\rm eff}}{T_{\rm sys}}= 391 \rm\ m^{2}K^{-1}$; bandwidth: $B=500$ MHz 

Integrated time: $\tau = 10$ minutes with NVSS observing model

\section{Estimate of the total observing time}

\begin{equation}
t=\frac{{\tau}A_{\rm sky}}{FoV}
\end{equation}

For SKA1\_SUR, field of view: $FoV =18$ \, (deg$^2$);
integrated time: $\tau = 10$ minutes with NVSS observing model; sky coverage: $A_{\rm sky}$ (equation A.3)
}
\end{appendices}

\end{multicols}
\end{document}